%% file: main.tex
\documentclass[10pt,conference]{IEEEtran}

\usepackage{cite}
\usepackage{amsmath,amssymb,amsfonts}
\usepackage{algorithmic}
\usepackage{graphicx}
\usepackage{textcomp}
\def\BibTeX{{\rm B\kern-.05em{\sc i\kern-.025em b}\kern-.08em
    T\kern-.1667em\lower.7ex\hbox{E}\kern-.125emX}}
\usepackage{booktabs}
\usepackage{xcolor}
\usepackage{epsfig}
\usepackage{epstopdf}
\usepackage{svg}
\graphicspath{{fig/}}
\usepackage{mathtools}
\usepackage{siunitx}
\usepackage{algorithmic}
\usepackage{algorithm}
\usepackage{tikzit}
\usepackage{enumitem}
\usepackage{lipsum}
\usepackage{subcaption}
\usepackage{steinmetz}
\usepackage{soul,color}
\usepackage{subcaption}  
\usepackage{bm}    

\usepackage[nolist,nohyperlinks]{acronym}

\usepackage{array} 
\newcolumntype{C}[1]{>{\centering\arraybackslash}p{#1}} 
\newcolumntype{L}[1]{>{\raggedright\arraybackslash}p{#1}} 

\input{styles_ICC_WM.tikzstyles}
\usepackage{amsthm}

\definecolor{applegreen}{rgb}{0.55, 0.71, 0.0}
\definecolor{awesome}{rgb}{1.0, 0.13, 0.32}
\definecolor{azure(colorwheel)}{rgb}{0.0, 0.5, 1.0}
\definecolor{darklavender}{rgb}{0.45, 0.31, 0.59}
\definecolor{cyan(process)}{rgb}{0.0, 0.72, 0.92}
\definecolor{brightmaroon}{rgb}{0.76, 0.13, 0.28}
\definecolor{ao(english)}{rgb}{0.0, 0.5, 0.0}
\definecolor{brightturquoise}{rgb}{0.03, 0.91, 0.87}
\definecolor{bondiblue}{rgb}{0.0, 0.58, 0.71}

\input{acronyms.tex}

\newcommand{\CLASSINPUTtoptextmargin}{1.8cm}

\usepackage{fancyhdr}
\fancypagestyle{acceptedpaper}{
  \fancyhf{} 
  \fancyhead[L]{\footnotesize ACCEPTED FOR PUBLICATION IN IEEE INTERNATIONAL CONFERENCE ON COMMUNICATIONS (ICC) 2026}
}

\begin{document}
\title{
Structured Latent Dynamics in Wireless CSI via Homomorphic World Models
}

\author{\IEEEauthorblockN{Salmane Naoumi\IEEEauthorrefmark{3}, Mehdi Bennis\IEEEauthorrefmark{2}, and Marwa Chafii\IEEEauthorrefmark{1}\IEEEauthorrefmark{3}}
	\IEEEauthorblockA{\IEEEauthorrefmark{1}Engineering Division, New York University (NYU), Abu Dhabi, UAE.}
\IEEEauthorblockA{\IEEEauthorrefmark{2}Center for Wireless Communications, University of Oulu, Finland.}
\IEEEauthorblockA{\IEEEauthorrefmark{3}NYU WIRELESS, NYU Tandon School of Engineering, New York, USA.}
}
\maketitle

\thispagestyle{acceptedpaper}

\IEEEpubid{}

\begin{abstract}
We introduce a self-supervised framework for learning predictive and structured representations of wireless channels by modeling the temporal evolution of channel state information (CSI) in a compact latent space. Our method casts the problem as a world modeling task and leverages the Joint Embedding Predictive Architecture (JEPA) to learn action-conditioned latent dynamics from CSI trajectories. To promote geometric consistency and compositionality, we parameterize transitions using homomorphic updates derived from Lie algebra, yielding a structured latent space that reflects spatial layout and user motion. Evaluations on the DICHASUS dataset show that our approach outperforms strong baselines in preserving topology and forecasting future embeddings across unseen environments. The resulting latent space enables metrically faithful channel charts, offering a scalable foundation for downstream applications such as mobility-aware scheduling, localization, and wireless scene understanding.
\end{abstract}


\section{Introduction}
\label{sec:introduction}
\input{sections/introduction}

\section{System Model}
\label{sec:system_model}
\input{sections/sys_model}


\section{Proposed solution}
\label{sec:methods}
\input{sections/methods}

\section{Numerical Evaluation}
\label{sec:numerical_eval}
\input{sections/results}

\section{Conclusion}
\label{sec:conclusion}
\input{sections/conclusion}

\section*{Acknowledgment}
%
This work was supported in part by the NYU Abu Dhabi Center for Artificial Intelligence and Robotics, funded by Tamkeen under the Research Institute Award CG010, and by the Technology Innovation Institute (TII). The authors also acknowledge use of the High Performance Computing resources at New York University Abu Dhabi.
\bibliographystyle{ieeetr}
\bibliography{references_ha}{}

\end{document}

%% file: styles_ICC_WM.tikzstyles
\tikzstyle{edge sharp}=[{|->}, dashed, thick]
\tikzstyle{csi transition}=[{|->}, thick]
\tikzstyle{phi cell}=[-, fill=none, draw={rgb,255: red,254; green,111; blue,94}]
\tikzstyle{new edge style 1}=[->, dashed, thick]
\tikzstyle{new edge style 2}=[{|-}, dashed]
\tikzstyle{exp cell}=[-, fill=none, draw={rgb,255: red,191; green,191; blue,191}]
\tikzstyle{new edge style 0}=[-]
\tikzstyle{new edge style 3}=[-, fill=white, tikzit fill=white, tikzit draw={rgb,255: red,191; green,191; blue,191}]
\tikzstyle{CSIs block}=[-, fill=none, tikzit draw={rgb,255: red,128; green,128; blue,128}, line width=0.25mm]
\tikzstyle{online encoder bloc}=[-, fill={rgb,255: red,254; green,111; blue,94}, draw=black, tikzit fill={rgb,255: red,254; green,111; blue,94}, tikzit draw=black]
\tikzstyle{vectorblock}=[-, fill={rgb,255: red,161; green,202; blue,241}, draw=black]
\tikzstyle{vectorEdge }=[-, draw=black]
\tikzstyle{new edge style 4}=[-, thick, dashed, dash pattern=on 2mm off 2mm]
\tikzstyle{new edge style 5}=[-, fill={rgb,255: red,0; green,191; blue,255}]
\tikzstyle{VectorEMA}=[->, thick]
\tikzstyle{Edgethick}=[-, thick]
\tikzstyle{new edge style 6}=[-, draw=black, fill={rgb,255: red,140; green,146; blue,172}]
\tikzstyle{ActionsBox}=[-, draw={rgb,255: red,140; green,146; blue,172}]
\tikzstyle{new edge style direct}=[-, dashed, thick]
\tikzstyle{LinkPlain}=[-, thick]
\tikzstyle{LossL1}=[draw={rgb,255: red,255; green,8; blue,0}, thick, <->]

%% file: acronyms.tex
\begin{acronym}
	\acro{ACDD}{Alamouti with cyclic delay diversity}
	\acro{URLLC}{ultra-reliable low-latency communications}
	\acro{3GPP}{third generation partnership project}
	\acro{PHY}{physical layer}
	\acro{MIMO}{multiple-input multiple-output}
	\acro{SIMO}{single-input multiple-output}
	\acro{MISO}{multiple-input single-output}
	\acro{SISO}{single-input single-output}
	\acro{MRC}{maximum-ratio combining}
	\acro{SNR}{signal-to-noise ratio}
	\acro{CP}{cyclic prefix}
	\acro{CDD}{cyclic delay diversity}
	\acro{FSC}{frequency-selective channel}
	\acro{STC}{space-time coding}
	\acro{FFT}{fast Fourier transform}
        \acro{PS}{phase shift}
	\acro{LMMSE}{linear minimum mean-squared error}
	\acro{FER}{frame error rate}
	\acro{OFDM}{orthogonal frequency division multiplexing}
	\acro{OCDM}{orthogonal chirp division multiplexing}
	\acro{FSC}{frequency-selective channel}
	\acro{CSI}{channel state information}
	\acro{LMMSE-PIC}{linear minimum mean squared error with parallel interference cancellation}
	\acro{PFE}{perfect-feedback equalizer}
	\acro{FD}{frequency domain}
	\acro{PDP}{power delay profile}
	\acro{PDF}{probability density function}
	\acro{DFT}{discrete Fourier transform}
	\acro{ICI}{inter-carrier interference}
	\acro{OTFS}{orthogonal time frequency space}
	\acro{AWGN}{additive white Gaussian noise}
	\acro{SWH}{sparse Walsh-Hadamard}
	\acro{LLR}{log-likelihood ratio}
	\acro{PMF}{probability mass function}
	\acro{CRC}{cyclic redundancy check}
	\acro{PAM}{pulse amplitude modulation}
	\acro{QAM}{quadrature amplitude modulation}
	\acro{FWHT}{fast Walsh-Hadamard transform}
	\acro{MAP}{maximum a-posteriori}
	\acro{SC}{single-carrier}
	\acro{ISI}{inter-symbol interference}
	\acro{ZP}{zero-padding}
	\acro{BCJR}{Bahl, Cocke, Jelinek, and Raviv}
	\acro{ISAC}{integrated sensing and communication}
	\acro{DFRC}{dual function radar communication}
	\acro{UE}{user equipment}
	\acro{PSD}{power spectral density}
	\acro{BS}{base station}
	\acro{HPA}{high power amplifier}
	\acro{MIMO}{multiple-input, multiple-output}
	\acro{PAPR}{peak-to-average power ratio}
	\acro{AWGN}{additive white Gaussian noise}
	\acro{KPI}{key performance indicator}
	\acro{KPIs}{key performance indicators}  
	\acro{D2D}{device-to-device}
	\acro{ISAC}{integrated sensing and communication}
	\acro{DFRC}{dual-functional radar and communication}
	\acro{AoA}{angle of arrival}
	\acro{ToA}{time of arrival}
	\acro{EVM}{error vector magnitude}
	\acro{CPI}{coherent processing interval}
	\acro{eMBB}{enhanced mobile broadband}
	\acro{URLLC}{ultra-reliable low latency communications}
	\acro{mMTC}{massive machine type communications}
	\acro{QCQP}{quadratic constrained quadratic programming}
	\acro{BnB}{branch and bound}
	\acro{SER}{symbol error rate}
	\acro{LFM}{linear frequency modulation}
	\acro{ADMM}{alternating direction method of multipliers}
	\acro{BS}{base station}
	\acro{LoS}{line-of-sight}
	\acro{CFO}{carrier frequency offset}
	\acro{UL}{uplink}
	\acro{DL}{downlink}
	\acro{ULA}{uniform linear antenna array}
	\acro{MOO}{multi-objective optimization}
	\acro{IBO}{input back-off}
	\acro{QPSK}{quadrature phase shift keying}
	\acro{CCDF}{complementary cumulative distribution function}
	\acro{OFDM}{orthogonal frequency-division multiplexing }
	\acro{IoT}{internet of things}
	\acro{CRB}{Cram\'er-Rao bound}
	\acro{OFDM}{orthogonal frequency-division multiplexing}
	\acro{DL}{downlink}
	\acro{i.i.d}{independently and identically distributed}
	\acro{UL}{uplink}
	\acro{LoS}{line of sight}
	\acro{DFT}{discrete Fourier transform}
	\acro{HPA}{high power amplifier}
	\acro{IBO}{input power back-off}
	\acro{MIMO}{multiple-input, multiple-output}
	\acro{PAPR}{peak-to-average power ratio}
	\acro{AWGN}{additive white Gaussian noise}
	\acro{KPI}{key performance indicator}
	\acro{KPIs}{key performance indicators}  
	\acro{D2D}{device-to-device}
	\acro{RSU}{roadside unit}
	\acro{ISAC}{integrated sensing and communication}
	\acro{DFRC}{dual-functional radar and communication}
	\acro{AoA}{angle of arrival}
	\acro{AoAs}{angles of arrival}
	\acro{AoD}{angle of departure}
	\acro{AoDs}{angles of departure}
	\acro{ToA}{time of arrival}
	\acro{ToAs}{times of arrival}
	\acro{EVM}{error vector magnitude}
	\acro{CPI}{coherent processing interval}
	\acro{eMBB}{enhanced mobile broadband}
	\acro{URLLC}{ultra-reliable low latency communications}
	\acro{mMTC}{massive machine type communications}
	\acro{QCQP}{quadratic constrained quadratic programming}
	\acro{BnB}{branch and bound}
	\acro{SER}{symbol error rate}
	\acro{LFM}{linear frequency modulation}
	\acro{ADMM}{alternating direction method of multipliers}
	\acro{CCDF}{complementary cumulative distribution function}
	\acro{MISO}{multiple-input, single-output}
	\acro{CSI}{channel state information}
	\acro{OTFS}{orthogonal time frequency space}
	\acro{LDPC}{low-density parity-check}
	\acro{BCC}{binary convolutional coding}
	\acro{IEEE}{Institute of Electrical and Electronics Engineers}
	\acro{ULA}{uniform linear antenna}
	\acro{B5G}{beyond 5G}
	\acro{MOOP}{multi-objective optimization problem}
	\acro{MLD}{maximum likelihood}
	\acro{ML}{machine learning}
	\acro{FML}{fused maximum likelihood}
	\acro{RF}{radio frequency}
	\acro{AM/AM}{amplitude modulation/amplitude modulation}
	\acro{AM/PM}{amplitude modulation/phase modulation}
	\acro{BS}{base station}
	\acro{MM}{majorization-minimization}
	\acro{LNCA}{$\ell$-norm cyclic algorithm}
	\acro{OCDM}{orthogonal chirp-division multiplexing}
	\acro{TR}{tone reservation}
	\acro{COCS}{consecutive ordered cyclic shifts}
	\acro{LS}{least squares}
	\acro{CVE}{coefficient of variation of envelopes}
	\acro{BSUM}{block successive upper-bound minimization}
	\acro{ICE}{iterative convex enhancement}
	\acro{SOA}{sequential optimization algorithm}
	\acro{BCD}{block coordinate descent}
	\acro{SINR}{signal to interference plus noise ratio}
	\acro{MICF}{modified iterative clipping and filtering}
	\acro{PSL}{peak side-lobe level}
	\acro{SDR}{semi-definite relaxation}
	\acro{KL}{Kullback-Leibler}
	\acro{ADSRP}{alternating direction sequential relaxation programming}
	\acro{MUSIC}{multiple signal classification}
	\acro{EVD}{eigenvalue decomposition}
	\acro{SVD}{singular value decomposition}
	\acro{MSE}{mean squared error}
	\acro{SNR}{signal-to-noise ratio}
	\acro{CRB}{Cram\'er-Rao bound}
	\acro{QPSK}{quadrature phase shift keying}
	\acro{RCS}{radar cross-section}
	\acro{NOMA}{non-orthogonal multiple access}
	\acro{IRS}{intelligent reflecting surface}
	\acro{IoT}{Internet of things}
	\acro{UAV}{unmanned aerial vehicle}
	\acro{V2X}{vehicle-to-everything}
	\acro{SDRs}{software-defined radios}
	\acro{FDD}{frequency-division duplex}
	\acro{NR}{new radio}
	\acro{FIM}{Fisher information matrix}
	\acro{PDF}{probability density function}    
	\acro{CRLB}{Cram\'er-Rao Lower Bound}
	\acro{CORDIC}{coordinate rotation digital computer}
	\acro{CIR}{channel impulse response}
	\acro{TO}{timing offset}
	\acro{FO}{frequency offset}
	\acro{CPI}{coherent processing interval}
	\acro{FFT}{fast Fourier transform}
	\acro{IFFT}{inverse fast Fourier transform}
	\acro{DFT}{discrete Fourier transform}
	\acro{DoA}{direction of arrival}
	\acro{MLP}{multi-layer perceptron}
	\acro{NN}{neural network}
	\acro{DL}{deep learning}
	\acro{MLE}{maximum likelihood estimator}
	\acro{TDD}{time-division duplexing}
	\acro{LWA}{leaky-wave antenna}
	\acro{mmWave}{milli-meter wave}
	\acro{C2VNN}{convolutional complex-valued neural network}
    \acro{SAGE}{space-alternating generalized expectation-maximization}
    \acro{ESPRIT}{estimation of signal parameters via rotational invariant techniques}
    \acro{CNN}{convolutional neural network}
    \acro{FNN}{feedforward neural network}
    \acro{GPU}{graphics processing unit}
    \acro{FLOPS}{floating point operations per second}
    \acro{FPGA}{field programmable gate array}
    \acro{SNN}{spiking neural network}
    \acro{FTO}{fractional timing offset}
    \acro{ppm}{parts Per million}
    \acro{MAE}{mean absolute error}
    \acro{GPS}{global positioning system}
    \acro{MMSE}{minimum mean squared error}
    \acro{RMSE}{root mean squared error}
    \acro{IDFT}{inverse discrete Fourier transform}
    \acro{CS}{compressive sensing}
    \acro{DML}{deterministic maximum likelihood}
    \acro{HPC}{high-performance computing}
    \acro{MAC}{multiply-accumulate operation}
    \acro{UPA}{uniform planar antenna array}
    \acro{MPC}{multipath components}
    \acro{AoD}{azimuth of departure}
    \acro{EoD}{elevation of departure}
    \acro{AoA}{azimuth of arrival}
    \acro{EoA}{elevation of arrival}
    \acro{EM}{expectation-maximization}
    \acro{SAGE}{space-alternating generalized EM}
    \acro{CNN}{convolutional neural network}
    \acro{3DCNN}{three-dimensional convolutional neural network}
    \acro{ReLU}{rectified linear unit}
    \acro{MHSA}{multi-head self attention}
    \acro{FLOP}{floating-point operation}
    \acro{6D}{six-dimensional}
    \acro{JEPA}{Joint-Embedding Predictive Architecture}
    \acro{AP}{access point}
    \acro{WM}{world model}
    \acro{EMA}{exponential moving average}
    \acro{MDP}{Markov Decision Process}
    \acro{IDM}{inverse dynamics modeling}
    \acro{FCF}{forward charting function}
    \acro{CT}{continuity}
    \acro{TW}{trustworthiness}
    \acro{KS}{kruskal’s stress}
    \acro{RD}{rajski’s distance}
    \acro{PCA}{principal component analysis}
    \acro{GRU}{gated recurrent unit}
    \acro{AI}{artificial intelligence}
    \acro{CC}{channel charting}
    \acro{FiLM}{feature-wise linear modulation} 
\end{acronym}

%% file: sections/introduction.tex
\Acp{WM} have emerged as a powerful paradigm in \ac{AI}, enabling agents to internalize environmental dynamics and simulate future states rather than react solely to immediate observations. At their core, \acp{WM} construct compact latent representations of the world and leverage them to predict the consequences of actions, reason about outcomes, and guide decision-making under uncertainty. Foundational works, such as that of Ha and Schmidhuber~\cite{ha2018worldmodels}, demonstrated how compressing high-dimensional sensory inputs into a latent space enables agents to plan and act more efficiently. More recent frameworks, including \acp{JEPA}~\cite{lecun2022path}, shift focus to predicting hidden or future representations directly in latent space, offering a self-supervised path toward learning causal and temporally coherent abstractions. Such models support internal simulations that enhance long-horizon reasoning, counterfactual inference, and policy planning~\cite{sora_wm, ding2025worldmodels}.

Integrating \acp{WM} into wireless communication systems offers a promising route toward autonomous and adaptive network intelligence~\cite{sun2025agiwireless}. Wireless environments pose unique challenges: observations (e.g., \ac{CSI}) are high-dimensional and noisy, channel dynamics are governed by complex physical processes, and real-time constraints demand low-latency, safe decision-making. Embedding a \ac{WM} within a wireless agent can help anticipate future channel states, evaluate strategies before deployment, and generalize to unseen conditions. Recent works such as Wireless Dreamer~\cite{zhao2025worldmodelscognitiveagents} and MobiWorld~\cite{chai2025mobiworldworldmodelsmobile} illustrate how world models embedded in edge agents enable simulation-based efficient planning, improved generalization, and robust policy optimization.

While prior efforts have focused on leveraging \acp{WM} for wireless control and decision-making, an open question remains: can the latent space of a \ac{WM} also serve as a meaningful representation of the radio environment? \Acf{CC} addresses a related goal by embedding high-dimensional \ac{CSI} into a lower dimensional latent space that preserves spatial proximity~\cite{studer2018channelcharting}. Classical approaches rely on handcrafted dissimilarity metrics derived from angular or delay domain features and apply manifold learning or Siamese contrastive training~\cite{adp_paper}. However, these methods are typically static, lack temporal awareness, and require domain-specific priors.

In this work, we build upon recent advances in predictive representation learning for wireless channels, particularly the \ac{JEPA}-based formulation of~\cite{wirelessJEPA}, which conditions on user velocity to forecast future channel embeddings. While effective, their method follows a two-stage curriculum, first pretraining a conventional \ac{CC} model and then learning dynamics atop that representation. By contrast, we propose a unified, self-supervised predictive \ac{WM} trained end-to-end from raw \ac{CSI} sequences without intermediate supervision. Our architecture follows the \ac{JEPA} formulation~\cite{jepa, pldm}, augmented with action conditioning and structured latent transitions inspired by recent advances in equivariant representation learning~\cite{hae}.
A key novelty lies in our introduction of homomorphic latent dynamics: instead of treating latent transitions as generic mappings, we enforce compositional structure by learning action representations that operate as smooth and consistent transformations in latent space. This approach enables the model to predict the outcomes of arbitrary action sequences with high geometric fidelity, without requiring prior knowledge of the transformation group or restrictions on the action space. As a result, our model constructs a predictive \ac{CC}, a latent geometry that not only captures spatial structure but also supports temporally coherent rollouts.

In summary, our contributions are as follows:
\begin{itemize}
    \item We propose a self-supervised predictive \ac{WM} for wireless channel modeling, trained on temporally coherent \ac{CSI} trajectories and equipped with action-conditioned latent transitions.
    
    \item We introduce a homomorphic transition operator based on Lie algebra parameterizations and matrix exponentials, enabling structured and compositional dynamics in the latent space. To the best of our knowledge, this is the first application of homomorphic latent \ac{JEPA}-style architectures to radio channel modeling.
    
    \item We show that the learned encoder functions as a \ac{FCF}, mapping \ac{CSI} into a metrically faithful latent space. We quantitatively evaluate the resulting predictive \ac{CC} using standard metrics such as \ac{TW}, \ac{CT}, \ac{KS}, and \ac{RD}.
    
    \item We conduct extensive experiments on the DICHASUS dataset, demonstrating that our model outperforms strong baselines including \ac{MLP}, \ac{GRU}, and \ac{FiLM} predictors, and generalizes effectively to unseen environments without retraining.
\end{itemize}

%% file: sections/sys_model.tex
We consider a wireless sensing setting in which a single-antenna \ac{UE} communicates with $B$ spatially distributed \acp{BS}, each equipped with an $M$-element antenna array. The \ac{UE} transmits an \ac{OFDM} waveform over \( N_{\tt{sub}} \) subcarriers. At each discrete time step \( t \), all \acp{BS} synchronously record the downlink \ac{CSI} in the frequency domain, resulting in a wideband multi-antenna tensor
\(
    \mathbf{H}^{(n)}_t \in \mathbb{C}^{B \times M \times N_{\tt{sub}}},
\)
where \( \mathbf{H}^{(n)}_t \) denotes the \ac{CSI} observed at time $t$ in trajectory $n$. Each CSI snapshot is paired with a timestamp \( t^{(n)}_t \in \mathbb{R} \) and a ground-truth user position \( \mathbf{x}^{(n)}_t \in \mathbb{R}^3 \), yielding a sequence of spatial-temporal observations
\[
    \mathcal{S} = \left\{ \left( \mathbf{H}^{(n)}_t, \mathbf{x}^{(n)}_t, t^{(n)}_t \right) \right\}_{t=0}^{T_n}, \quad n = 1, \dots, N,
\]
where \( T_n \) is the number of snapshots in the $n^{\text{th}}$ trajectory. To model the temporal dynamics of channel evolution, we augment each trajectory with control inputs reflecting the \ac{UE}'s motion. Specifically, we define the control vector \( \mathbf{a}^{(n)}_t \in \mathbb{R}^d \) as a $2$D velocity signal estimated from consecutive positions. This yields a temporally ordered trajectory with control annotations of the form
\begin{equation}
\label{eq:trajectories_dataset}
    \mathcal{D} = \left\{ \tau^{(n)} = \left( \mathbf{H}^{(n)}_0, \mathbf{a}^{(n)}_0, \mathbf{H}^{(n)}_1, \dots, \mathbf{a}^{(n)}_{T_n - 1}, \mathbf{H}^{(n)}_{T_n} \right) \right\}_{n=1}^N.
\end{equation}
The velocity based inputs \( \mathbf{a}^{(n)}_t \) act as action-like signals governing state transitions, allowing us to interpret the dataset as a sequence of interactions from an underlying dynamical system.

This structure lends itself naturally to a \ac{MDP} abstraction, defined as \( \mathcal{M}=(\mathcal{S}, \mathcal{A}, \mu, p) \), where \( \mathcal{S} \) is the state space of \ac{CSI} observations, \( \mathcal{A} \subset \mathbb{R}^d \) is the control (or action) space, \( \mu \) is the initial state distribution, and \( p(\mathbf{H}_{t+1} \mid \mathbf{H}_t, \mathbf{a}_t) \) defines the stochastic transition dynamics driven by motion.
Although no reward signals are defined in this setup, the \ac{MDP} structure enables the learning of predictive models of channel evolution without explicit supervision, thereby supporting a wide range of downstream wireless applications such as beam tracking, mobility-aware scheduling, and proactive resource management. As detailed in Section~\ref{sec:methods}, we propose to learn such models via structured latent representations, without relying on spatial labels or handcrafted similarity metrics.

%% file: sections/methods.tex
\begin{figure*}[t]
  \centering
  \includegraphics[width=\textwidth,pagebox=artbox]{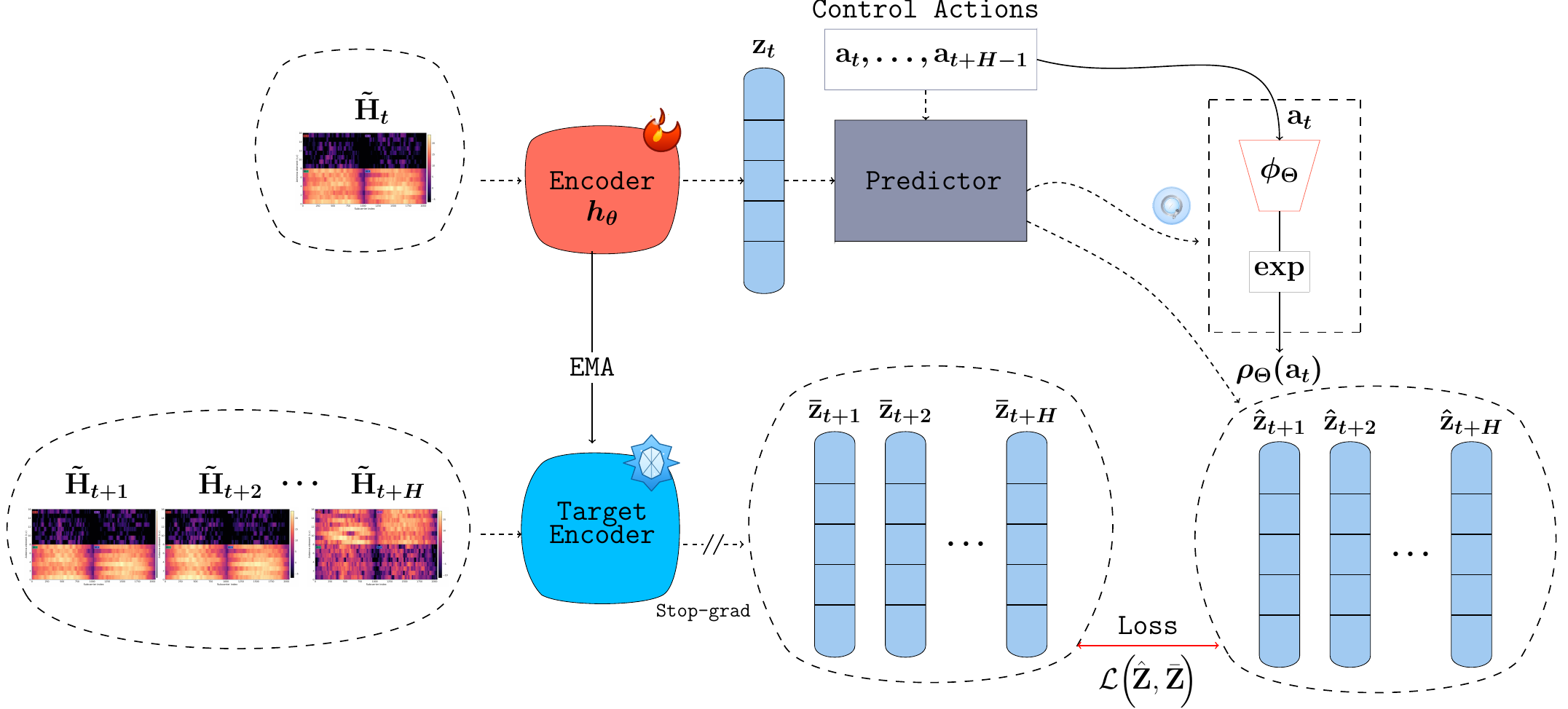}
  \caption{Overview of the proposed \ac{WM} architecture. The model encodes real-valued \ac{CSI} tensors into latent vectors, then evolves them through a structured, Lie group-based transition model conditioned on user motion. It is trained via self-supervised rollout prediction and regularized to preserve geometric consistency and action-awareness.}
\label{fig:model_architecture}
\end{figure*}
In contrast to traditional manifold learning and metric-based \ac{CC} techniques that rely on pairwise dissimilarities, we propose a structured \ac{WM} framework to learn predictive, geometry-aware latent representations of multi-antenna wideband \ac{CSI}. As illustrated in Figure~\ref{fig:model_architecture}, we treat the evolution of the wireless channel as a latent dynamical system driven by user motion, modeled as control-conditioned transitions in latent space.
%
%
We assume that the wireless environment remains predominantly static over the duration of each trajectory, i.e., the spatial layout of reflectors and scatterers is fixed and no abrupt occlusions occur. Hence, we model the dynamics of the observed channel as primarily governed by user motion. While small-scale effects and noise may contribute to variability, we assume these manifest as temporally smooth and locally continuous deviations. Accordingly, we take the \ac{UE}'s $2$D velocity as the primary control signal driving state transitions.
Under this abstraction, the model learns a compact latent representation \( \mathbf{z}_t = \zeta(\tilde{\mathbf{H}}_t) \in \mathbb{R}^D \) from a preprocessed \ac{CSI} tensor \( \tilde{\mathbf{H}}_t \), and a transition operator \( \psi \) such that
\begin{equation}
\label{eq:rollout_prediction}
\hat{\mathbf{z}}_{t+h} = \psi(\mathbf{z}_t; \mathbf{a}_t, \dots, \mathbf{a}_{t+h-1}),
\end{equation}
enabling structured prediction of future embeddings based on sequences of the \ac{UE} motion.

Each complex \ac{CSI} observation \( \mathbf{H}^{(n)}_t \in \mathbb{C}^{B \times M \times N_{\tt{sub}}} \) is transformed into a real-valued tensor \( \tilde{\mathbf{H}}^{(n)}_t \in \mathbb{R}^{2 \times (B \cdot M) \times N_{\tt{sub}}} \) by concatenating magnitude and phase. These are processed by a shared encoder \( \zeta = h_\theta \), yielding latent embeddings \( \mathbf{z}^{(n)}_t = h_\theta(\tilde{\mathbf{H}}^{(n)}_t) \in \mathbb{R}^D \). During training, we extract sub-sequences of length \( H+1 \) from each trajectory, along with \( H \) associated control vectors
\begin{equation}
\tau^{(n)}_t = \left( \tilde{\mathbf{H}}^{(n)}_t, \mathbf{a}^{(n)}_t, \tilde{\mathbf{H}}^{(n)}_{t+1}, \dots, \mathbf{a}^{(n)}_{t+H-1}, \tilde{\mathbf{H}}^{(n)}_{t+H} \right).
\end{equation}

In predictive representation learning, prior work such as PLDM~\cite{pldm} models latent dynamics using a generic transition function \( f_\Theta \) implemented as a generic \ac{NN}
\begin{equation}
    \hat{\mathbf{z}}_{t+h} = f_\Theta(\mathbf{z}_t, \mathbf{a}_t, \dots, \mathbf{a}_{t+h-1}).
\end{equation}
However, such mappings are often unstructured, limiting generalization across varying motion patterns. Instead, we propose a structured dynamics model inspired by~\cite{hae}, where control inputs induce transformations in the latent space via compositions of Lie group actions.
Specifically, we define a learnable mapping \( \phi_\Theta: \mathbb{R}^d \rightarrow \mathfrak{gl}(D) \) that projects control inputs \( \mathbf{a}_t \) into a Lie algebra. The corresponding Lie group element is then obtained via the matrix exponential
\begin{equation}
\rho_\Theta(\mathbf{a}_t) = \exp(\phi_\Theta(\mathbf{a}_t)).
\end{equation}
Future latents are then predicted via successive group compositions acting on the current state
\begin{equation}
\label{eq:rollout_eq_pred}
\hat{\mathbf{z}}_{t+h} = \left( \prod_{i=0}^{h-1} \rho_\Theta(\mathbf{a}_{t+i}) \right) \mathbf{z}_t.
\end{equation}
This homomorphic structure ensures temporal consistency, algebraic compositionality, and geometric coherence in the latent dynamics.

We train our model using a self-supervised predictive objective that encourages consistent latent rollouts across temporally aligned channel observations. Drawing inspiration from \acp{JEPA}~\cite{jepa,pldm}, the training framework adopts a teacher-student setup wherein the model is optimized to forecast future representations under action-conditioned latent dynamics. The encoder \( h_\theta \) generates online embeddings \( \mathbf{z}_t \), while a non-trainable target encoder \( h_{\text{EMA}} \) is maintained as an \ac{EMA} of the online encoder and produces reference embeddings \( \bar{\mathbf{z}}_{t+h} = h_{\text{EMA}}(\tilde{\mathbf{H}}_{t+h}) \) for each future step \( t + h \). The predicted latent \( \hat{\mathbf{z}}_{t+h} \) is computed via Eq.~\eqref{eq:rollout_eq_pred} using the online encoder and control sequence.

To train the latent predictor, we extract rollout segments of horizon \( H \) from each trajectory, organizing them into batches of size \( K \). Let \( \hat{\mathbf{Z}}, \bar{\mathbf{Z}} \in \mathbb{R}^{H \times K \times D} \) represent the predicted and target latent sequences, respectively. We begin by minimizing a teacher-forcing loss, aligning each predicted latent to its corresponding target using an \( \ell_1 \) loss
\begin{equation}
    \mathcal{L}_{\text{TF}} = \frac{1}{H K} \sum_{t=1}^H \sum_{b=1}^K \left\| \hat{\mathbf{Z}}_{t,b} - \bar{\mathbf{Z}}_{t,b} \right\|_1.
\end{equation}
To promote consistency over the full prediction horizon, we include an additional rollout loss that supervises only the terminal latent
\begin{equation}
    \mathcal{L}_{\text{roll}} = \frac{1}{K} \sum_{b=1}^K \left\| \hat{\mathbf{Z}}_{H,b} - \bar{\mathbf{Z}}_{H,b} \right\|_1.
\end{equation}

While these losses encourage local alignment, they do not explicitly regularize the structure of the latent space. To address this, we incorporate a VICReg-style regularization~\cite{vicreg}, which encourages non-degenerate and decorrelated embeddings. The variance term ensures that each latent dimension maintains sufficient spread across the batch
\begin{equation}
    \mathcal{L}_{\text{var}} = \frac{1}{H D} \sum_{t=1}^H \sum_{j=1}^D \max \left( 0, \gamma - \sqrt{\text{Var}(\bar{\mathbf{Z}}_{t,:,j}) + \epsilon} \right),
\end{equation}
\noindent where \(\gamma\) is a target standard deviation hyperparameter and \(\epsilon\) is a small constant introduced to ensure numerical stability. Additionally, we apply a covariance penalty that reduces linear correlations among embedding dimensions using the empirical covariance matrix.
\begin{equation}
    C(\bar{\mathbf{Z}}_t) = \frac{1}{K - 1} (\bar{\mathbf{Z}}_t - \bar{\bar{\mathbf{Z}}}_t)^\top (\bar{\mathbf{Z}}_t - \bar{\bar{\mathbf{Z}}}_t),
\end{equation}
\begin{equation}
    \mathcal{L}_{\text{cov}} = \frac{1}{H D} \sum_{t=1}^H \sum_{i \neq j} \left[ C(\bar{\mathbf{Z}}_t) \right]^2_{i,j},
\end{equation}
where \( \bar{\bar{\mathbf{Z}}}_t = \frac{1}{K} \sum_{b=1}^K \bar{\mathbf{Z}}_{t,b} \) denotes the batch mean at time \( t \).

To further reinforce the action-aware structure of the learned embeddings, we introduce an \ac{IDM} loss~\cite{IDM}. A frozen \ac{MLP} \( g_\psi \), initialized at the beginning of training, is tasked with decoding the control input from consecutive latent pairs. This encourages the learned representations to contain information predictive of motion
\begin{equation}
    \mathcal{L}_{\text{IDM}} = \frac{1}{H K} \sum_{t=1}^H \sum_{b=1}^K \left\| \mathbf{a}_{t-1,b} - g_\psi(\bar{\mathbf{Z}}_{t-1,b}, \bar{\mathbf{Z}}_{t,b}) \right\|_2^2.
\end{equation}

The final training objective aggregates all losses into a weighted sum
\begin{equation}
    \mathcal{L}_{\text{total}} = \lambda_{\text{TF}} \mathcal{L}_{\text{TF}} 
    + \lambda_{\text{roll}} \mathcal{L}_{\text{roll}} 
    + \lambda_{\text{var}} \mathcal{L}_{\text{var}} 
    + \lambda_{\text{cov}} \mathcal{L}_{\text{cov}} 
    + \lambda_{\text{IDM}} \mathcal{L}_{\text{IDM}},
\end{equation}
where we use default values \( \lambda_{\text{TF}} = 1.0 \), 
\( \lambda_{\text{roll}} = 2.0 \), 
\( \lambda_{\text{var}} = 2.0 \), 
\( \lambda_{\text{cov}} = 10.0 \), and 
\( \lambda_{\text{IDM}} = 1.0 \). 
These weights were selected via hyperparameter search on a held-out validation set and fixed across all experiments, unless stated otherwise.

Model training is conducted for $100$ epochs using the AdamW optimizer. The learning rate follows a cosine decay schedule with a $5\%$ linear warm-up, starting at \( 10^{-4} \), peaking at \( 3 \times 10^{-4} \), and decaying to \( 10^{-6} \). Weight decay is linearly ramped from \( 0.04 \) to \( 0.4 \) across training. A batch size of $128$ is used throughout, and the \ac{EMA} target encoder is updated with a decay factor of $0.9995$ to ensure smooth representation drift. This training setup enables the model to learn temporally predictive, geometrically structured, and action-consistent embeddings that support robust rollout forecasting and trajectory unfolding in wireless environments.

%% file: sections/results.tex
\begin{figure*}[!ht]
\centering

\begin{subfigure}[b]{0.24\textwidth}
    \centering
    \includegraphics[scale=0.11]{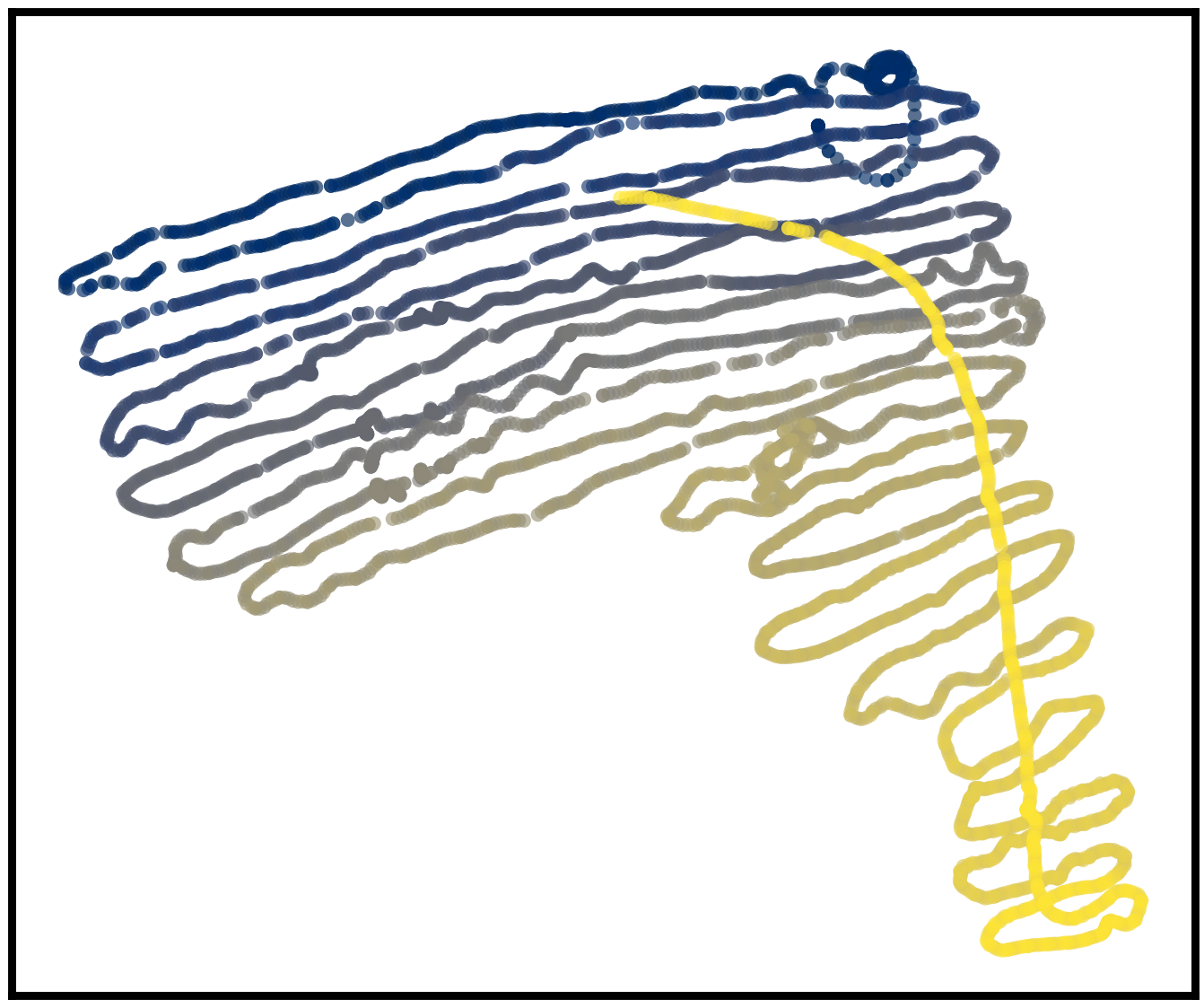}
    \caption{\texttt{cf02}: Ground truth}
\end{subfigure}
\hfill
\begin{subfigure}[b]{0.24\textwidth}
    \centering
    \includegraphics[scale=0.115]{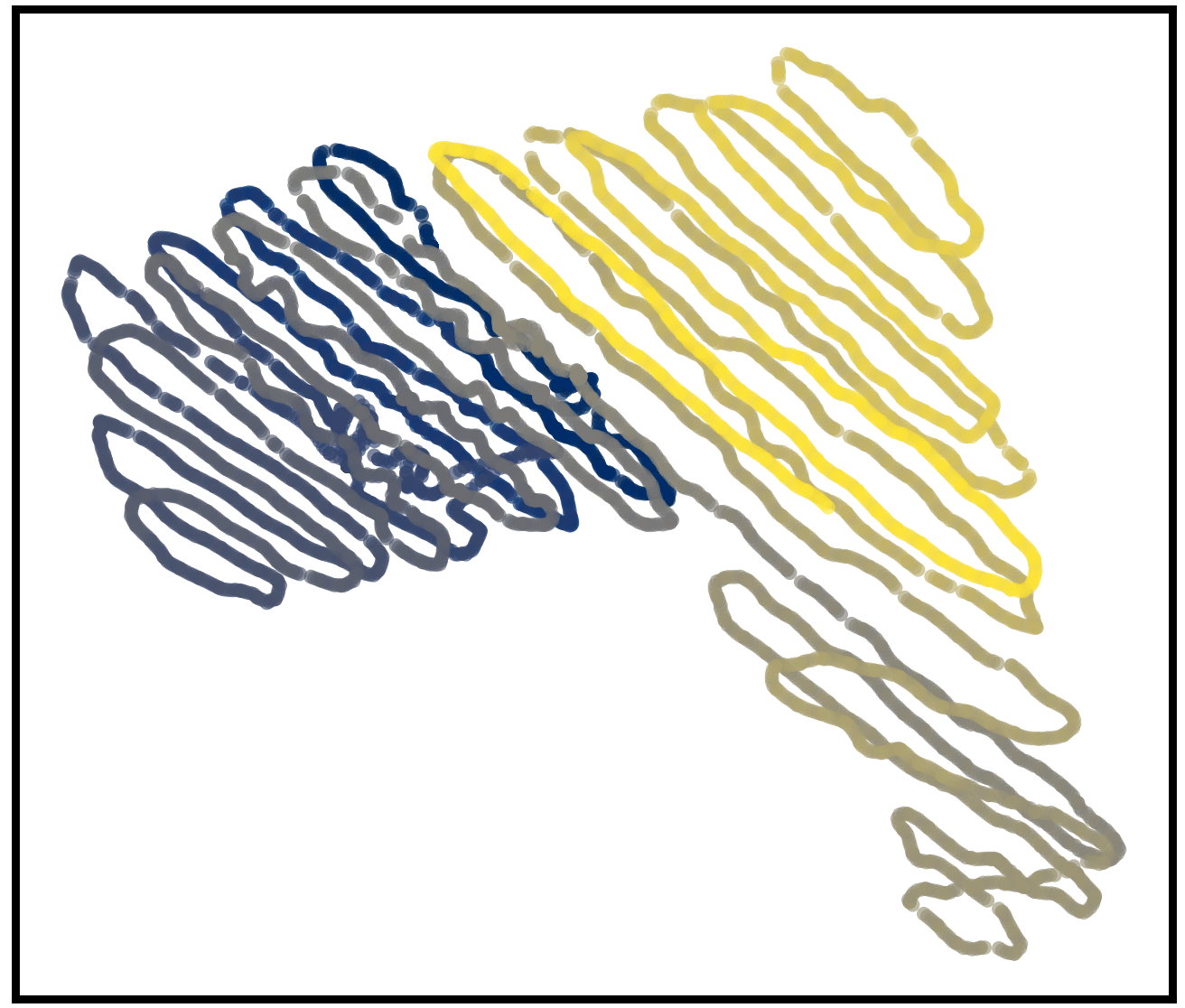}
    \caption{\texttt{cf03}: Ground truth}
\end{subfigure}
\hfill
\begin{subfigure}[b]{0.24\textwidth}
    \centering
    \includegraphics[scale=0.113]{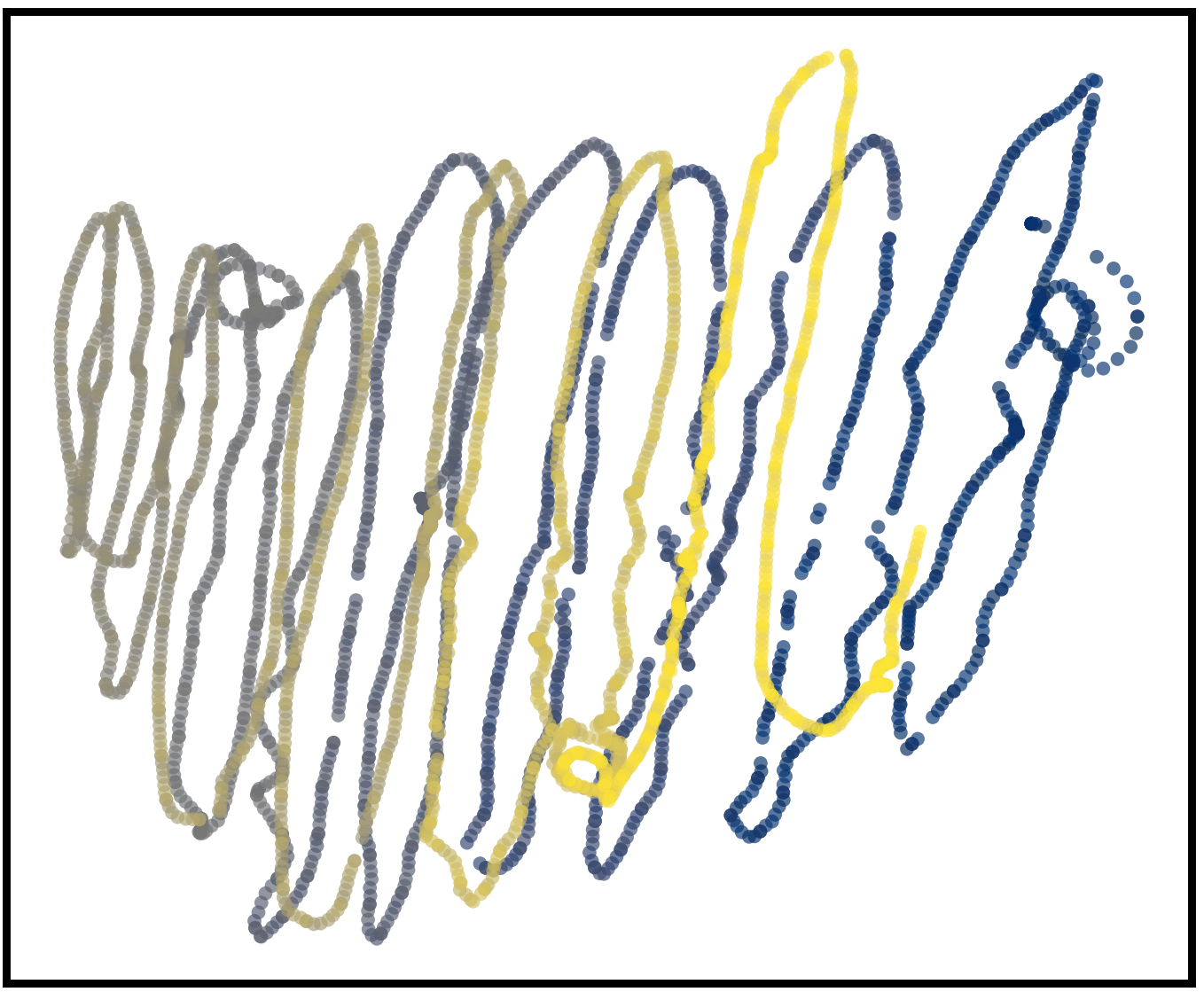}
    \caption{\texttt{cf06}: Ground truth}
\end{subfigure}
\hfill
\begin{subfigure}[b]{0.24\textwidth}
    \centering
    \includegraphics[scale=0.115]{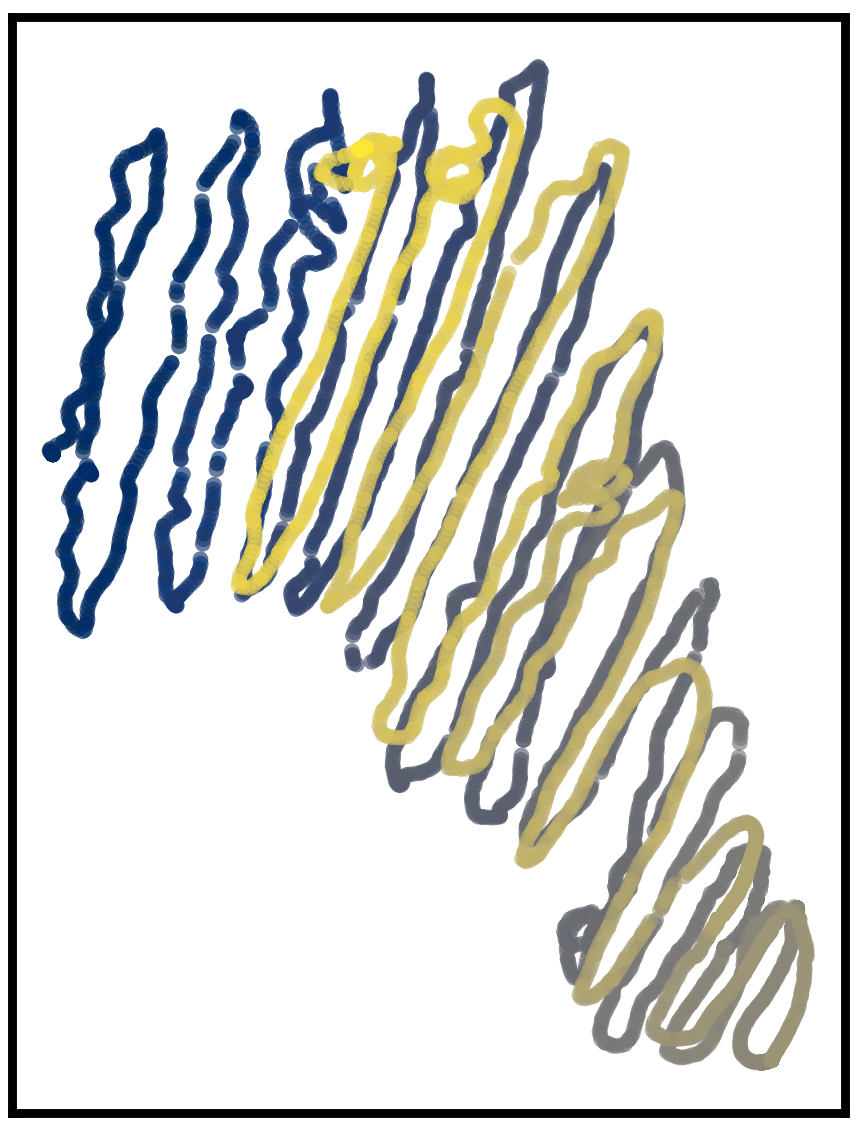}
    \caption{\texttt{cf05}: Ground truth}
\end{subfigure}

\vspace{1em}

\begin{subfigure}[b]{0.24\textwidth}
    \centering
    \includegraphics[scale=0.1]{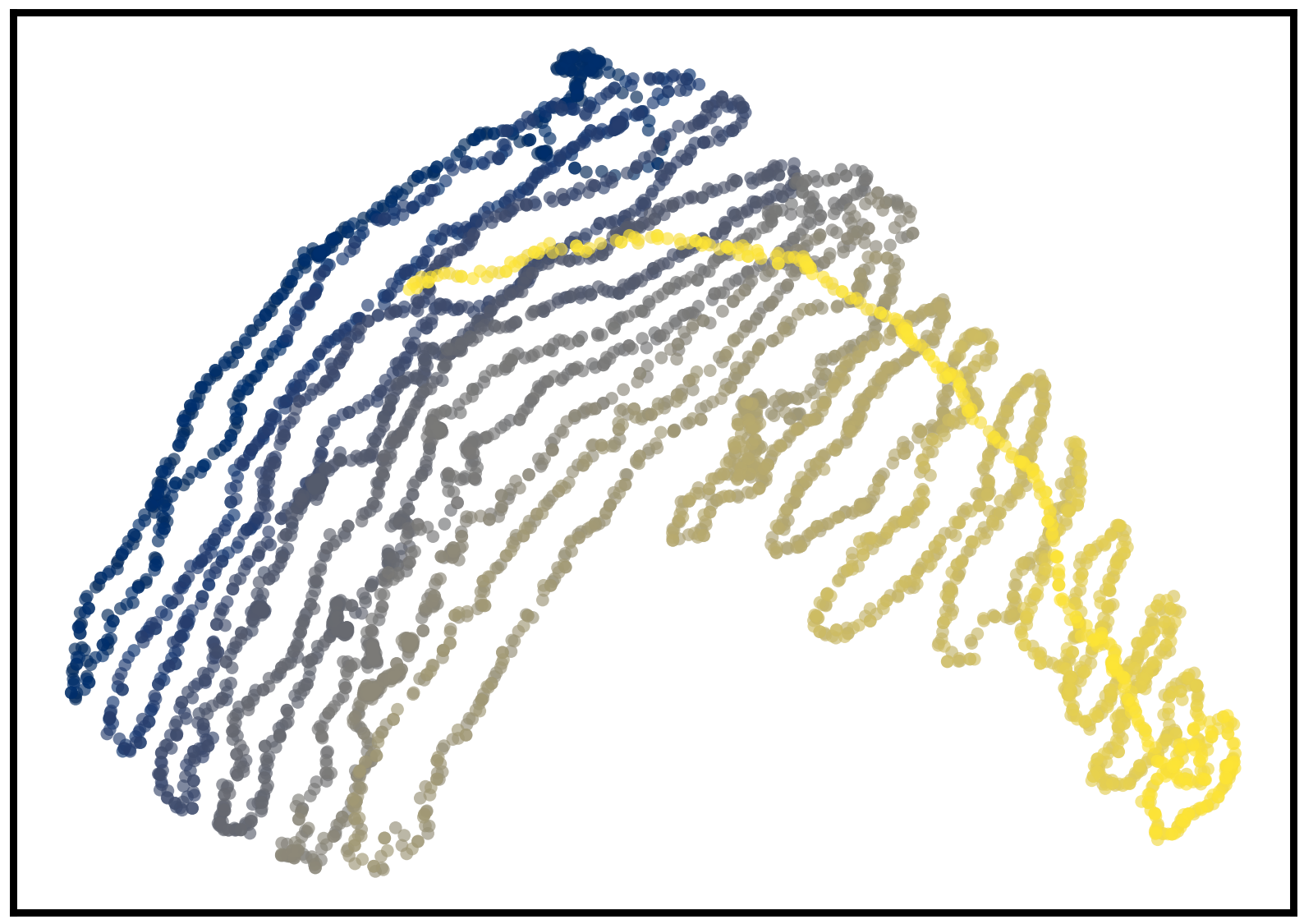}
    \caption{\texttt{cf02}: PCA rollout}
\end{subfigure}
\hfill
\begin{subfigure}[b]{0.24\textwidth}
    \centering
    \includegraphics[scale=0.1]{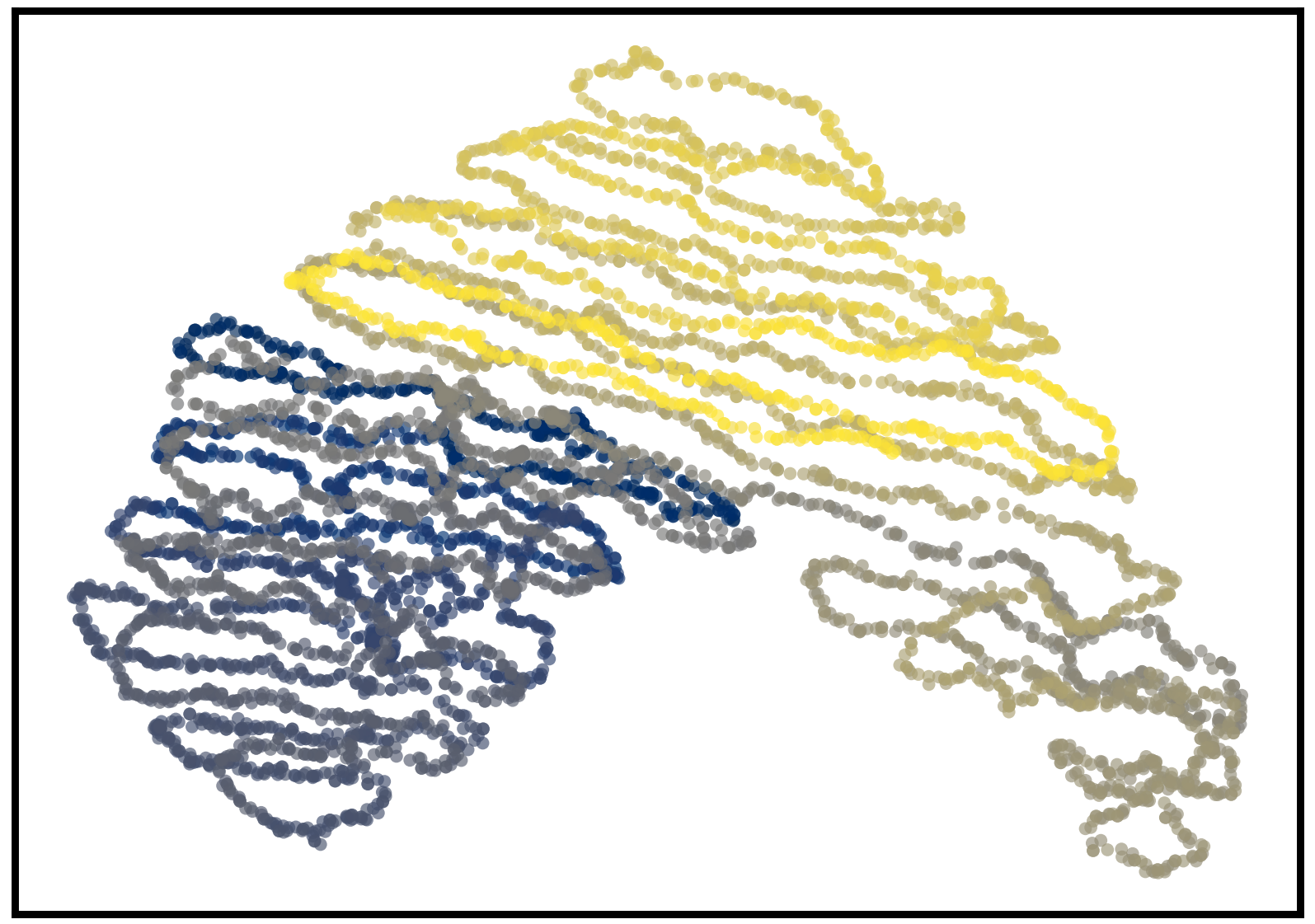}
    \caption{\texttt{cf03}: PCA rollout}
\end{subfigure}
\hfill
\begin{subfigure}[b]{0.24\textwidth}
    \centering
    \includegraphics[scale=0.1]{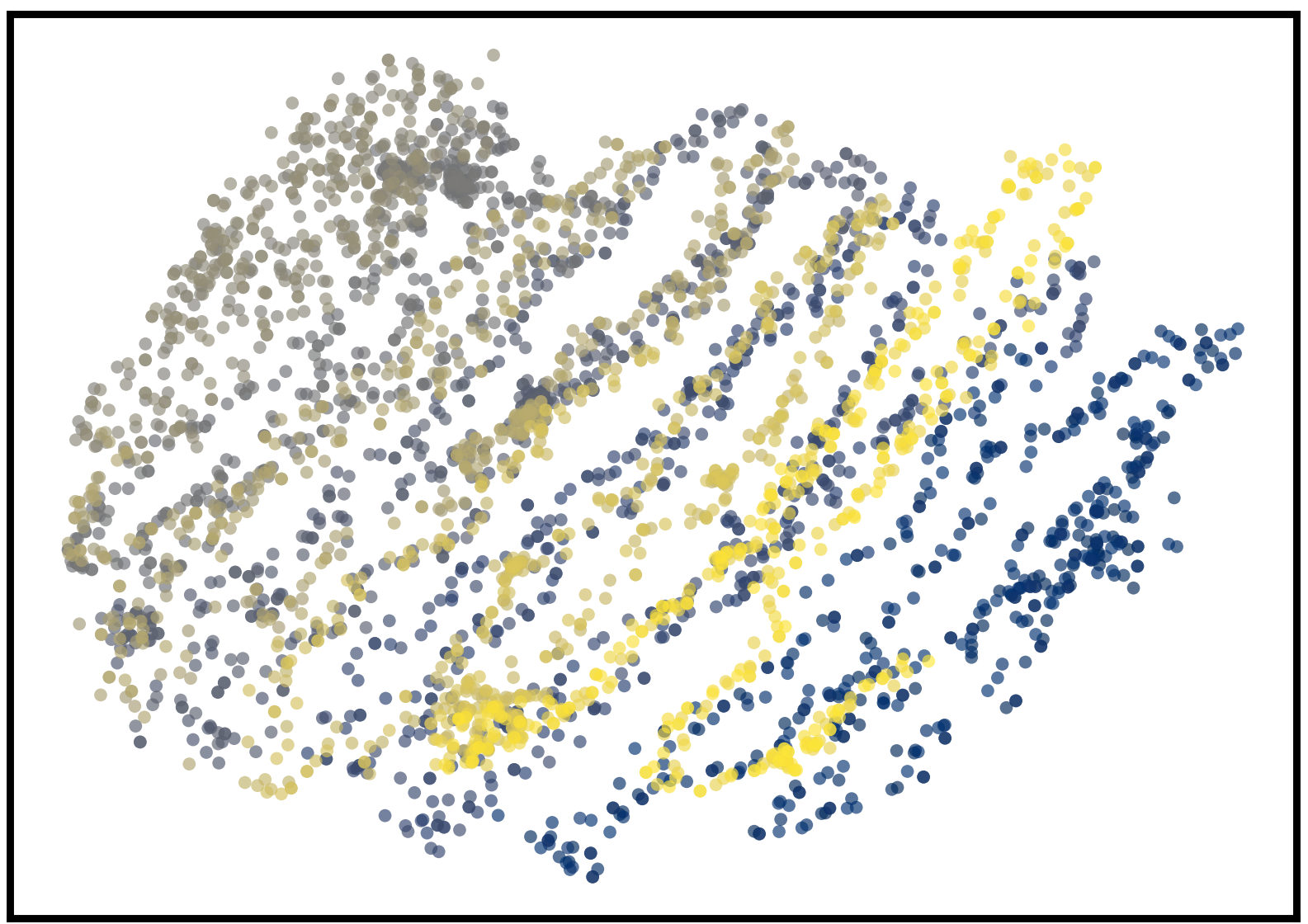}
    \caption{\texttt{cf06}: PCA rollout}
\end{subfigure}
\hfill
\begin{subfigure}[b]{0.24\textwidth}
    \centering
    \includegraphics[scale=0.1]{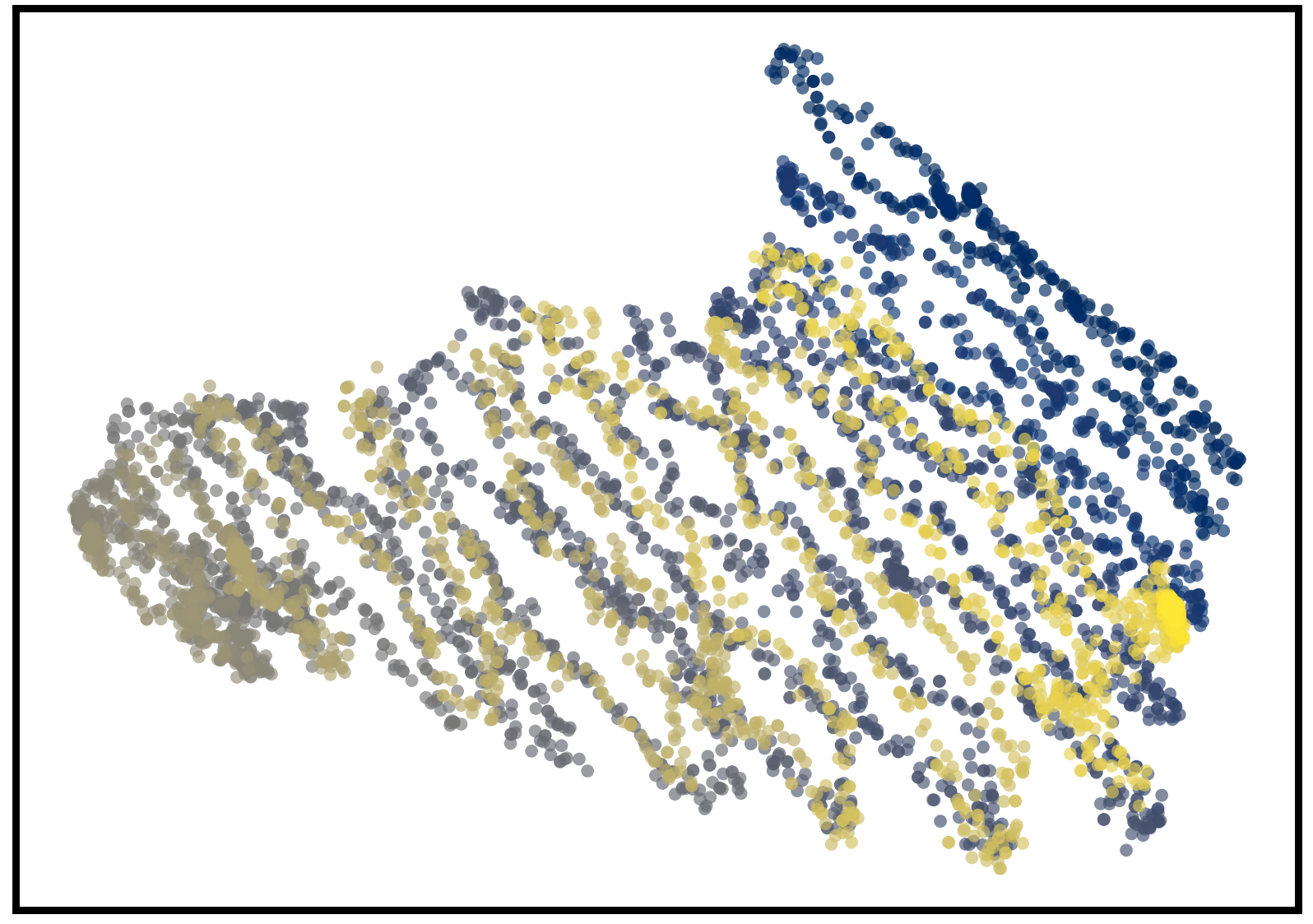}
    \caption{\texttt{cf05}: PCA rollout}
\end{subfigure}

\vspace{1em}

\begin{subfigure}[b]{0.24\textwidth}
    \centering
    \includegraphics[scale=0.1]{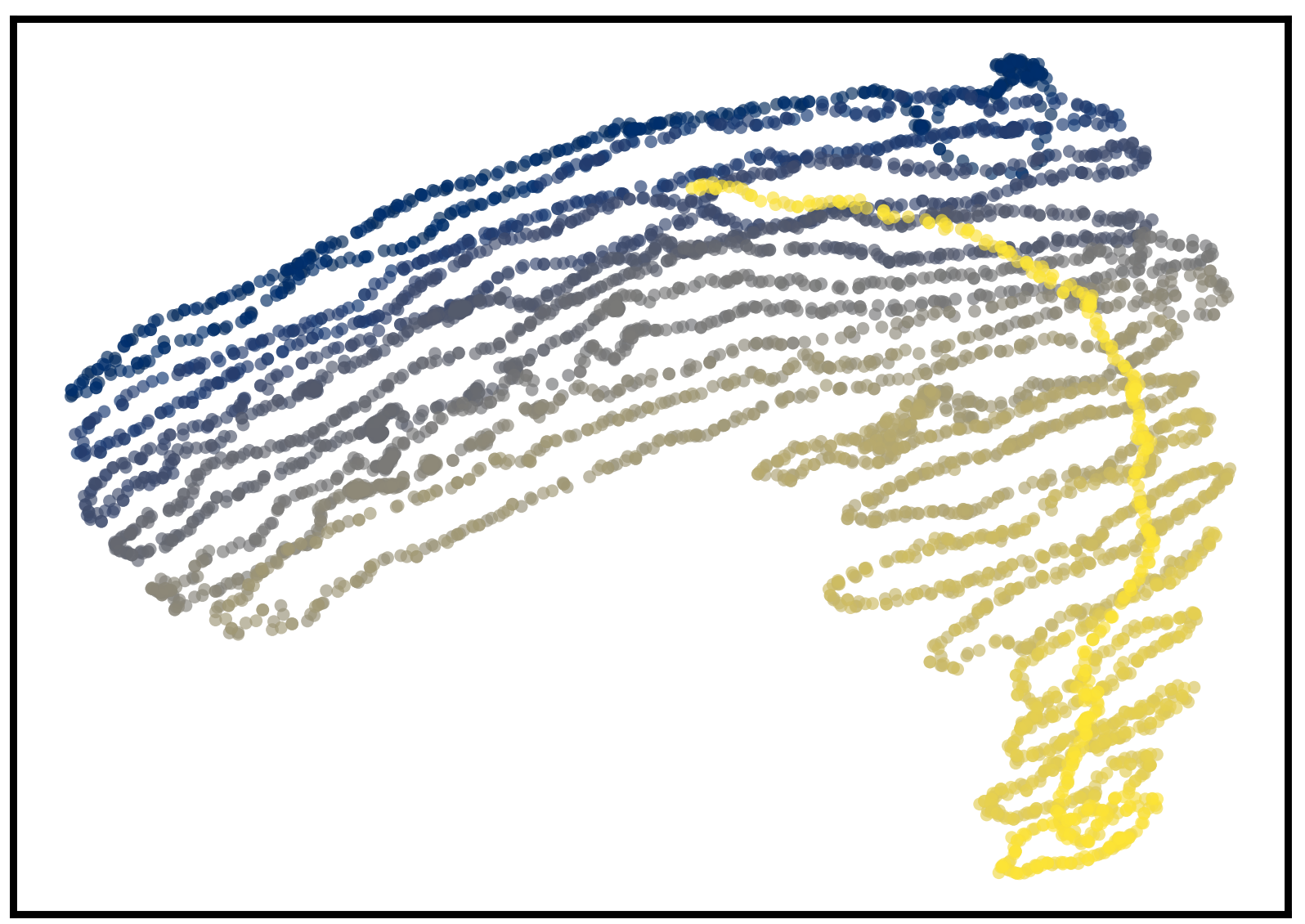}
    \caption{\texttt{cf02}: Aligned embedding}
\end{subfigure}
\hfill
\begin{subfigure}[b]{0.24\textwidth}
    \centering
    \includegraphics[scale=0.1]{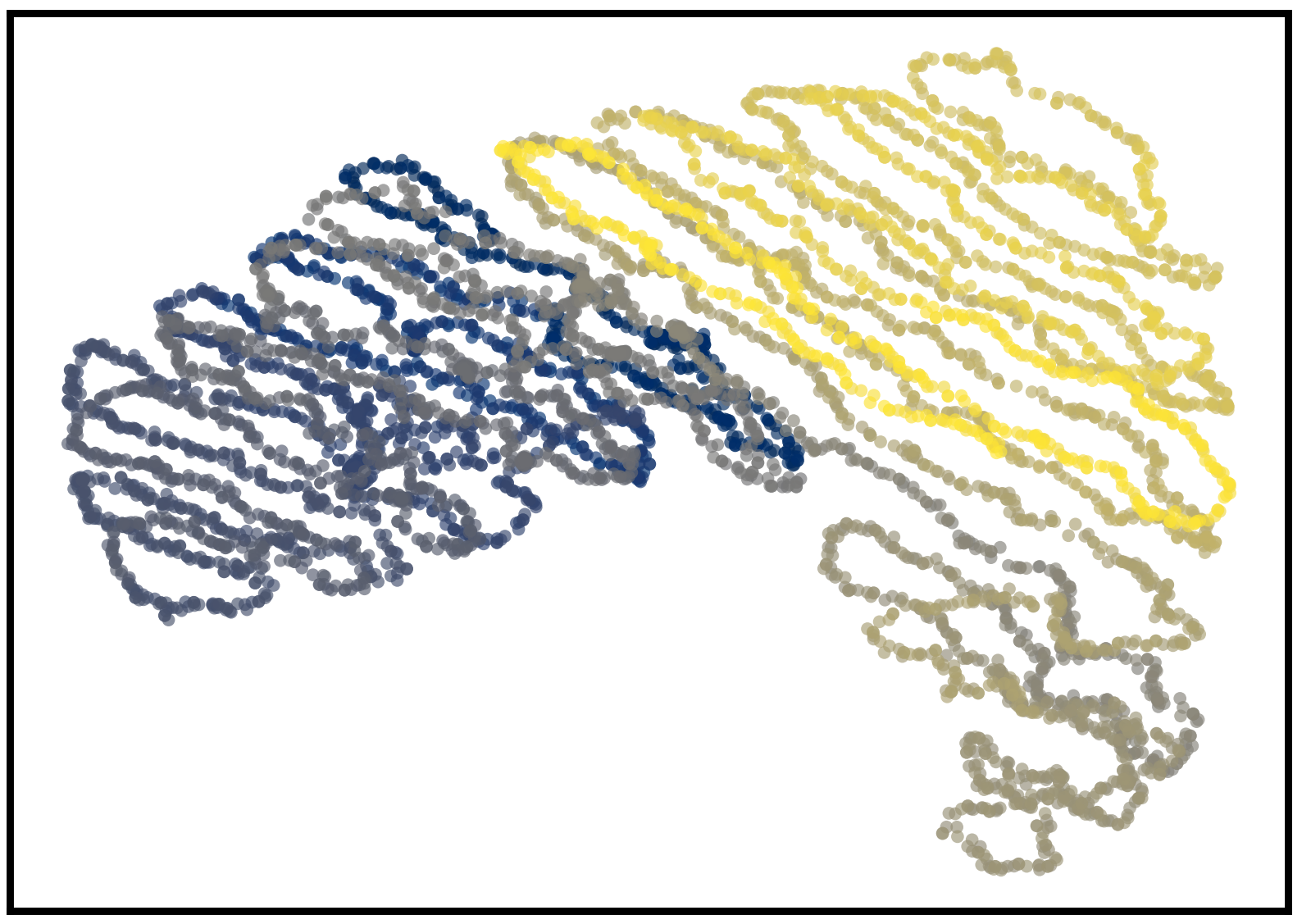}
    \caption{\texttt{cf03}: Aligned embedding}
\end{subfigure}
\hfill
\begin{subfigure}[b]{0.24\textwidth}
    \centering
    \includegraphics[scale=0.1]{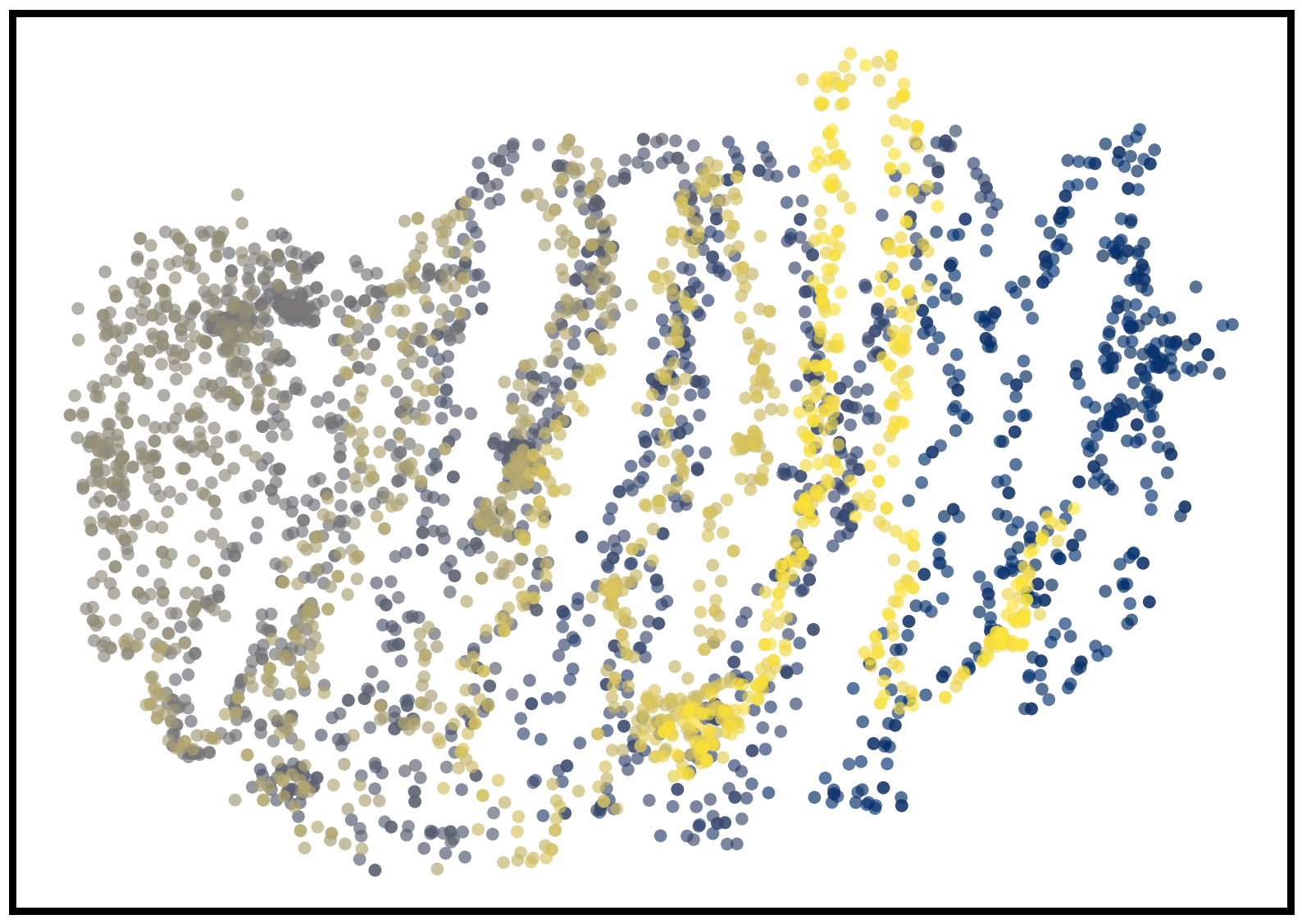}
    \caption{\texttt{cf06}: Aligned embedding}
\end{subfigure}
\hfill
\begin{subfigure}[b]{0.24\textwidth}
    \centering
    \includegraphics[scale=0.1]{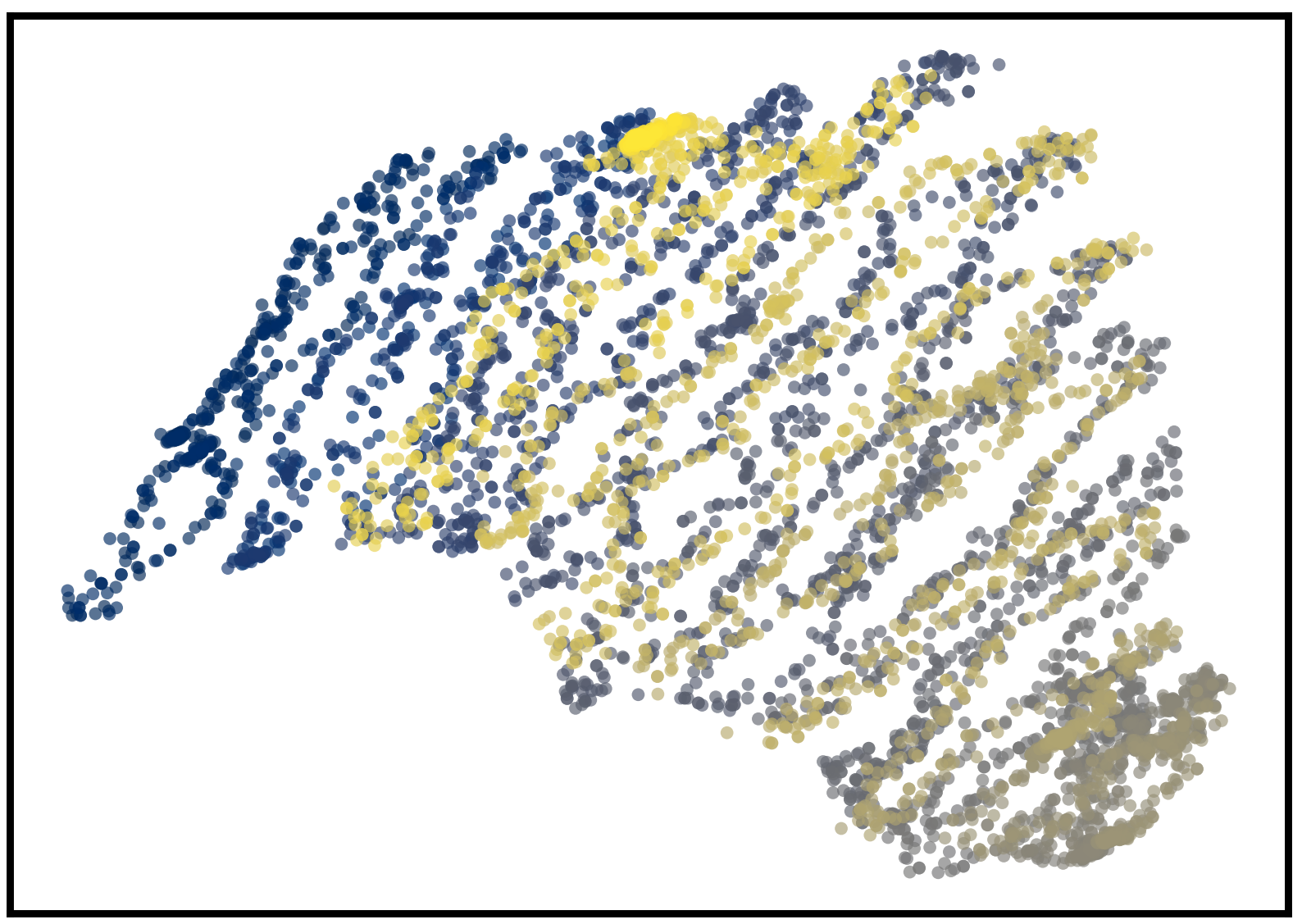}
    \caption{\texttt{cf05}: Aligned embedding}
\end{subfigure}

\caption{
\textbf{Visualization of latent rollouts vs. ground truth trajectories.} Each column corresponds to a scene (\texttt{cf02}, \texttt{cf03}, \texttt{cf06}, \texttt{cf05}). Top row: Ground truth user trajectories in physical space. Middle row: PCA projections of predicted latent rollouts using the trained model and input velocity sequences. Bottom row: Same latent rollouts, aligned (via Procrustes) to the ground-truth positions. The learned latent space exhibits spatial and temporal consistency across both seen and unseen environments.
}
\label{fig:chart_rollout}
\end{figure*}
\begin{table*}[!ht]
\centering
\normalsize
\setlength{\tabcolsep}{2pt}
\begin{tabular}{C{2.8cm}|C{3.5cm}|C{3.5cm}|C{3.5cm}|C{3.5cm}}
\toprule
\textbf{Predictor} & $\uparrow\,\text{TW}$ & $\uparrow\,\text{CT}$ & $\downarrow\,\text{KS}$ & $\downarrow\,\text{RD}$ \\
\midrule
Proposed & $\mathbf{0.9948}\,\pm\,0.0026$ & $\mathbf{0.9764}\,\pm\,0.0178$ & $\mathbf{0.1001}\,\pm\,0.0351$ & $\mathbf{0.7473}\,\pm\,0.0758$ \\
FiLM     & $0.9935\,\pm\,0.0033$ & $0.9640\,\pm\,0.0273$ & $0.1330\,\pm\,0.0488$ & $0.8149\,\pm\,0.0680$ \\
MLP      & $0.9927\,\pm\,0.0029$ & $0.9622\,\pm\,0.0168$ & $0.1234\,\pm\,0.0354$ & $0.7828\,\pm\,0.0667$ \\
GRU      & $0.9916\,\pm\,0.0039$ & $0.9680\,\pm\,0.0170$ & $0.1190\,\pm\,0.0340$ & $0.7818\,\pm\,0.0571$ \\
\bottomrule
\end{tabular}
\caption{
\textbf{Chart quality metrics on held-out subsets (\texttt{cf04}, \texttt{cf05}, \texttt{cf06}).} Mean $\pm$ standard deviation across trajectories. Higher is better for \ac{TW}, \ac{CT}; lower is better for \ac{KS}, \ac{RD}.
}
\label{tab:chart_metrics}
\end{table*}
We evaluate our approach on the DICHASUS dataset~\cite{dataset_dichasus}, which contains wideband indoor channel measurements acquired by a distributed channel sounder. The setup consists of a single antenna transmitter mounted on a mobile robot and $B = 4$ spatially distributed receiver arrays, each equipped with $M = 8$ antennas arranged in a $2 \times 4$ configuration, for a total of $32$ receive antennas. The system operates over $N_{\tt{sub}} = 1024$ \ac{OFDM} subcarriers spanning a $\SI{50}{\mega\hertz}$ bandwidth centered at $\SI{1.272}{\giga\hertz}$, with \ac{CSI} snapshots recorded every $\SI{40}{\milli\second}$. Each sample includes a complex valued frequency domain \ac{CSI} tensor $\mathbf{H} \in \mathbb{C}^{B \times M \times N_{\tt{sub}}}$ alongside ground truth positions and timestamps.
To reduce input dimensionality while retaining essential channel structure, we apply a time domain truncation that compresses the number of subcarriers to $64$. The resulting tensor is cast into a two channel real-valued representation containing the magnitude and wrapped phase of the complex coefficients, yielding inputs of shape $(2, 32, 64)$.
We use subsets \texttt{cf02}, \texttt{cf03}, and \texttt{cf07} for training, totaling $25{,}149$ samples, and evaluate on \texttt{cf04}, \texttt{cf05}, and \texttt{cf06}. While all subsets are recorded in the same L-shaped industrial space, they contain distinct motion paths, allowing us to assess generalization to novel trajectories. The \ac{CSI} samples are segmented into temporally coherent trajectories, as described in Section~\ref{sec:methods}, for sequential modeling.

The encoder \(h_\theta\) is a ResNet-style convolutional network~\cite{resnet}, with four stages of depths $[3, 4, 6, 3]$ and channel dimensions $[128, 192, 256, 256]$, each followed by downsampling. GeLU activations and BatchNorm are used throughout. The output is pooled and normalized to obtain a \( D = 128 \)-dimensional latent embedding, which we found to offer the best balance between training stability and charting quality across validation subsets. The dynamics model \(\phi_\Theta\) is a two-layer \ac{MLP} with GeLU activations, mapping control inputs to $D \times D$ matrix generators, which are exponentiated to yield linear operators in the latent space. The \ac{IDM} module \(g_\psi\) is a frozen two-layer \ac{MLP} with ReLU activations. We use a rollout horizon $H = 6$ and tube masking ratio $r_m = 0.15$.

The proposed architecture serves a predictive \ac{WM} trained on temporally coherent \ac{CSI} trajectories. We posit that the encoder, once trained, functions as a \ac{FCF}, projecting high-dimensional \ac{CSI} observations into a compact latent space wherein euclidean distances correlate with spatial displacements between the \ac{UE} positions. To quantify this behavior, we report four metrics from the channel charting literature: \ac{CT}, \ac{TW}, \Ac{KS}, and \Ac{RD}. \ac{CT} and \ac{TW} are optimal at $1$ and measure local structure preservation, while \ac{KS} and \ac{RD} are optimal at $0$ and reflect the global topological alignment.

Figure~\ref{fig:chart_rollout} visualizes predicted latent rollouts on representative training and test trajectories. Starting from an initial \ac{CSI} input, the model generates future embeddings conditioned solely on velocity sequences, without access to future \acp{CSI}. The resulting latent trajectories, visualized using \ac{PCA}, preserve both temporal coherence and geometric structure, closely matching the ground truth positions even in unseen environments. The third row shows aligned projections via Procrustes analysis for spatial interpretability.

Table~\ref{tab:chart_metrics} compares the proposed homomorphic transition model to baseline predictors trained under the same losses. The \ac{MLP} baseline models dynamics as an unconstrained function of $(z_t, a_t)$. FiLM~\cite{perez2018film} introduces action-dependent affine modulation
\begin{equation}
    z_{t+1} = z_t + \gamma(a_t) \odot z_t + \beta(a_t),
\end{equation}
where $\gamma(a_t), \beta(a_t) \in \mathbb{R}^D$ are learned via \acp{MLP}. The \ac{GRU} predictor adopts a recurrent architecture as in~\cite{wirelessJEPA}. Our method enforces compositional latent transitions via homomorphic mappings $\exp(\phi_\Theta(a_t))$, promoting algebraic structure and smoothness across action sequences.
%
%
The proposed method consistently outperforms all baselines across metrics, demonstrating that enforcing homomorphic latent transitions yields more structured and metrically faithful embeddings. Notably, our model generalizes to test trajectories never seen during training, highlighting the benefit of compositional structure for predictive and generalizable channel charting.

\begin{table}[!ht]
\centering
\setlength{\tabcolsep}{6pt}
\begin{tabular}{lcccc}
\toprule
\textbf{Loss Components} & \textbf{TW}$\uparrow$ & \textbf{CT}$\uparrow$ & \textbf{KS}$\downarrow$ & \textbf{RD}$\downarrow$ \\
\midrule
TF + Roll + VICReg + IDM     & \textbf{0.9996} & \textbf{0.9994} & \textbf{0.0388} & \textbf{0.6057} \\
TF + Roll + IDM & 0.9988 & 0.9934 & 0.0629 & 0.6605 \\
TF + Roll + VICReg           & 0.9984 & 0.9986 & 0.0640 & 0.7326 \\
TF + Roll                    & 0.9760 & 0.6443 & 0.5593 & 0.9719 \\
\bottomrule
\end{tabular}
\caption{
\textbf{Ablation of loss components on latent chart quality.} Results are reported for the testing sets using standard metrics. Adding VICReg and \ac{IDM} regularization significantly improves both local and global structure.
}
\label{tab:loss_ablation}
\end{table}
To evaluate the contribution of the individual loss components, we conduct an ablation study summarized in Table~\ref{tab:loss_ablation}. Removing both VICReg and \ac{IDM} regularization (Teacher-forcing + Rollout loss only) leads to a pronounced drop in performance, with reduced local structure (\ac{CT} = 0.6443) and degraded global alignment (\ac{RD} = 0.9719). Adding VICReg alone improves both \ac{TW} and \ac{CT}, demonstrating the importance of embedding regularization. Replacing VICReg with a predictive loss maintains moderate structure but yields higher distortion. The best performance is achieved when both VICReg and \ac{IDM} are included, suggesting that variance decorrelation and motion-aware supervision jointly encourage temporally stable and metrically faithful representations. These findings highlight the complementary role of information-theoretic and dynamics-aware objectives in structuring the latent space.

Finally, while our evaluation primarily focuses on geometric fidelity and rollout consistency of the learned latent space, we note that downstream applications such as beam selection, handover prediction, or user scheduling can be naturally built atop the proposed representation. However, such tasks typically require additional domain-specific components, labels, or control policies, which fall outside the core scope of this work. Our goal is to establish the predictive and compositional structure of the latent dynamics, which we consider a necessary precursor to reliable downstream reasoning. As a first step, we assume a predominantly static environment where channel evolution is governed by user motion. Extending this framework to handle dynamic settings with time-varying scatterers, non-line-of-sight conditions, or multipath variability is critical for real-world deployment. Moreover, incorporating constraints on computational efficiency such as latency, memory footprint, or FLOPs will be essential for enabling integration into latency-sensitive wireless pipelines. We leave both robustness and system-level optimization as promising directions for future work.

%% file: sections/conclusion.tex
In this work, we proposed a predictive world modeling framework tailored to wireless environments, leveraging the temporal structure of evolving \ac{CSI} measurements. By recasting channel charting as a latent dynamics learning problem, we introduced a self-supervised architecture based on \ac{JEPA} that learns action-conditioned transitions in a compact latent space. To enforce geometric consistency and compositionality, we incorporated homomorphic updates via Lie algebraic parameterizations, yielding a structured transition model capable of stable and interpretable rollouts. Our experiments on the DICHASUS dataset demonstrate that the learned latent space preserves spatial topology and generalizes across diverse motion trajectories. These findings suggest that predictive \acp{WM} offer a principled foundation for constructing geometry-aware wireless representations. Future work will investigate the integration of this approach into end-to-end systems for downstream tasks such as localization, adaptive beamforming, and autonomous wireless decision-making.